\documentclass[article,aps,pra%,changes,superscriptaddress,12pt,
citeautoscript,amsmath,amssymb,floatfix,
,bbm, verbatim,amsfonts,nofootinbib]{revtex4-1}
\usepackage{fancyhdr}
\fancypagestyle{titlepagestyle}{
	\fancyhf{} % Clear all header and footer fields
\fancyfoot[C]{Honourable Mention in the Gravity Research Foundation 2023 Awards for Essays on Gravitation.\\ This is more a gathering of arguments, rather than a novel contribution. Part of this work reprises\cite{oppenheim2018post}}

}
\pdfoutput=1
\usepackage{amsmath,bbold,bm}
\usepackage{amssymb}
\usepackage{epsfig}
\usepackage{xcolor}
\usepackage{amsthm}
\usepackage{appendix}
\usepackage{cancel}
\usepackage{stmaryrd}
\usepackage{soul}
\usepackage{ulem}
\usepackage{hyperref}
\usepackage[disable]{todonotes}
\usepackage{multirow}
\usepackage{hyperref}
\usepackage{xifthen}

\def\<{\langle}
\def\>{\rangle}

\newcommand{\be}{\begin{eqnarray} \begin{aligned}}
\newcommand{\ee}{\end{aligned} \end{eqnarray} }
\newcommand{\benn}{\begin{eqnarray*} \begin{aligned}}
		\newcommand{\eenn}{\end{aligned} \end{eqnarray*} }
\newcommand{\ben}{\begin{eqnarray} \begin{aligned}}
\newcommand{\een}{\end{aligned} \end{eqnarray} }

\newcommand{\bc}{\begin{center}}
	\newcommand{\ec}{\end{center}}

\newcommand{\beq}{\begin{eqnarray} \begin{aligned}}
\newcommand{\eeq}{\end{aligned} \end{eqnarray} }
\newcommand{\bea}{\begin{array}}
	\newcommand{\eea}{\end{array}}
\newcommand{\bee}{\begin{enumerate}}
	\newcommand{\eee}{\end{enumerate}}
\newcommand{\bei}{\begin{itemize}}
	\newcommand{\eei}{\end{itemize}}

\usepackage{amsfonts}

\def\01{\{0,1\}}

\newcommand{\ket}[1]{|#1\rangle}

\renewcommand{\sout}[1]{\ignorespaces}
\newcommand{\jono}[1]{\textcolor{black}{#1}}

\def\<{\langle}
\def\>{\rangle}

\newtheorem*{rep@theorem}{\rep@title}
\newcommand{\newreptheorem}[2]{
	\newenvironment{rep#1}[1]{
		\def\rep@title{#2 \ref{##1} (restatement)}
		\begin{rep@theorem}}
		{\end{rep@theorem}}}
\makeatother
\newreptheorem{thm}{Theorem}
\newreptheorem{lem}{Lemma}

 \DeclareMathAlphabet\mathbfcal{OMS}{cmsy}{b}{n}

\def\z{{z}}

\def\0mom{{\rate^{\alpha\beta}(\z)}}
\def\1mom{{\rate^{\alpha\beta}_1(\z)}}
\def\2mom{{\rate^{\alpha\beta}_2(\z)}}

\def\rate{{W}}

\renewcommand{\varrho}{\hat{\rho}}

\def\t0{0}

\begin{document}
%\frontmatter
%\include{reply_to_refs}
%\mainmatter
%

\thispagestyle{empty}

\title{Is it time to rethink quantum gravity?}
% "It's very difficult to find a black cat in a dark room, especially if there is no cat."

\author{Jonathan Oppenheim}

\affiliation{Department of Physics and Astronomy, University College London, Gower Street, London WC1E 6BT, United Kingdom {j.oppenheim@ucl.ac.uk}}
\date{March 31, 2023, midnight AoE}

\begin{abstract}
%Does the Pope have lips?
%It's well past due.
%An argument for why we should seriously consider the possibility that gravity may not be a quantum field.
Although it's widely believed that gravity should have a quantum nature like every other force, 
%While it may appear that gravity should have a quantum nature like every other force, 
the conceptual obstacles to constructing a quantum theory of gravity 
compel us to explore other perspectives. Gravity is not like any other force. It alone defines a universal space-time geometry, upon which quantum fields evolve. We feel gravity because matter causes space-time to bend. Time flows at unequal rates at different locations. 
% the bending of space-time, and
% Time flows at different rates at different points. 
%Gravity is intimately linked with  the unequal flow of time at different points.
The rate at which time flows, and the causal structure it provides, may be required to have a classical description in order for quantum theory to be well-formulated. I discuss arguments for this proposition, but ultimately conclude that we must turn to experiment to guide us. 
%Fortunately, such experiments can already place constraints on the alternative.
% may become feasible in the coming decades,  some which 
\end{abstract}
%relations that gravity provides, may require a classical 

%\let\clearpage\relax
\maketitle
\thispagestyle{titlepagestyle}

\clearpage
\pagestyle{plain}
\pagenumbering{arabic}

If we were to choose a well-timed 
%arbitrary 
event to mark the beginning of 
the {\it General Relativity Renaissance}\cite{renaissance},
%\nocite{will1989renaissance,kaiser2018price,rickles2020covered,blum2020renaissance,goenner2017golden} 
why not pick Bryce DeWitt writing
''New Directions for Research in the Theory for Gravitation"\cite{dewitt1953new} 
which won  the 1953 Gravity Research Foundation essay prize. 
Written seventy years ago, DeWitt forcefully made the case for pursuing a quantum theory of gravity, a program that had begun more than twenty years earlier by Rosenfeld\cite{rosenfeld1930quantelung,rosenfeld1930gravitationswirkungen}.
But when I revisit his essay today, it is his reflection on the prevailing views of his era that strikes me more than the arguments he musters in favour of his position:

 \begin{quote} [O]ne may well ask to know the reasons for attempting quantization of the gravitational field in the first place. As a matter of fact, the overwhelming weight of opinion of physicists is opposed to the attempt... It may actually be that the gravitational field is the one and only field which is {\it not} quantized in Nature. The gravitational field, with its attendant phenomena, could, under these circumstances, constitute the ultimate classical level which must be postulated, even in the quantum theory, in order to have a consistent ''quantum theory of measurement.'' \end{quote}
Even DeWitt was not completely convinced when it comes to the question of quantising or not quantising the gravitational field, concluding that
''at the present stage of the game, there is little to choose between the two possibilities.''

\begin{figure}[h]
	\includegraphics[width=.7\textwidth]{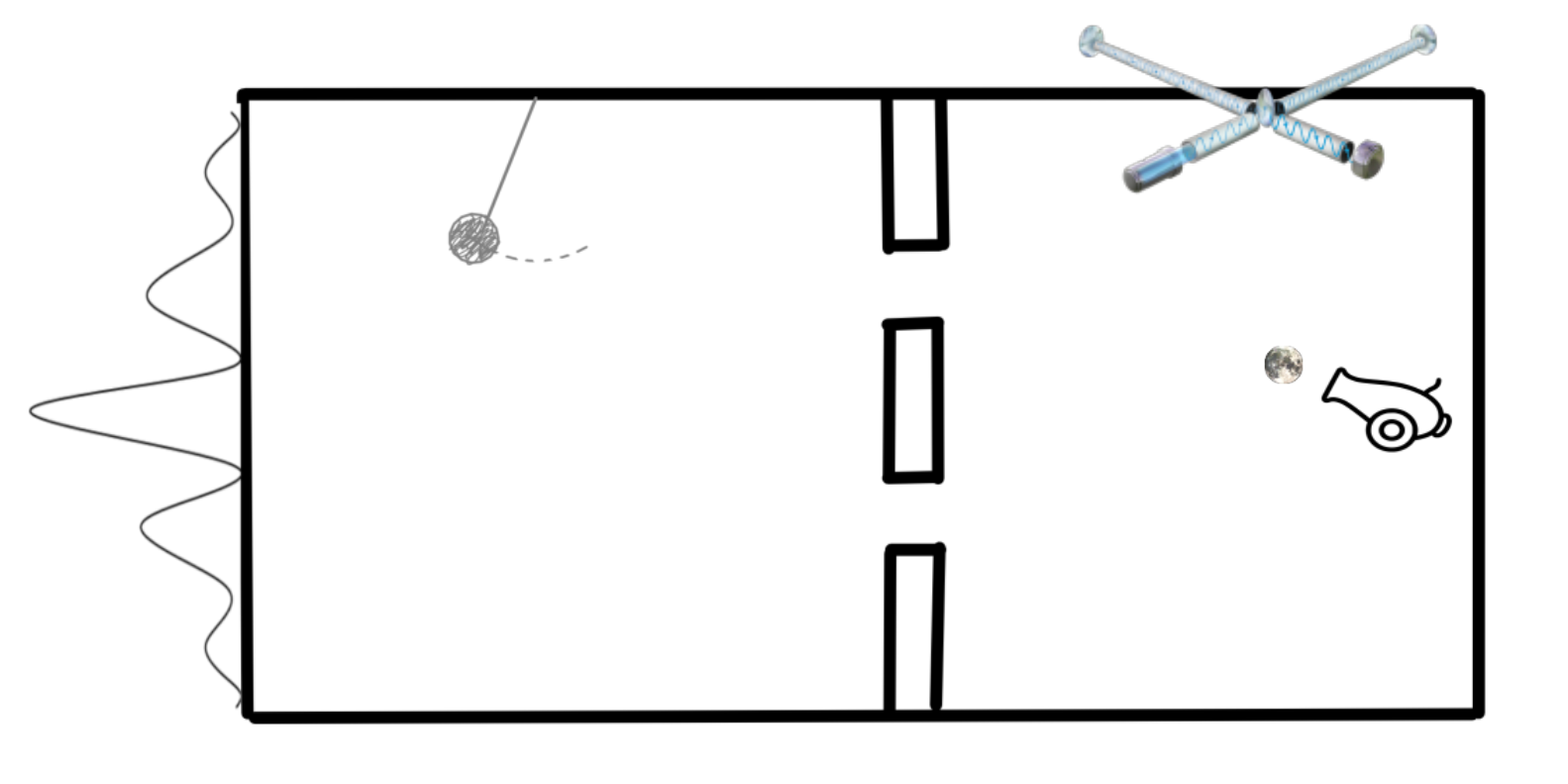}%{doubleslit.png}%
	\caption{A variation of the {\it gedanken experiment} proposed by Feynman\cite{cecile2011role,Feynman:1996kb-note} and Aharonov\cite{AharonovParadoxes-note}. A massive particle going through the left slit is in state $\ket{L}$ and produces a different gravitational field, then if it had gone through the slit on the right when it is in state $\ket{R}$. The particle  produces a different gravitational field depending on whether it is going through the left or right slit. If the gravitational field is classical, it can be measured to arbitrary accuracy (using a pendulum and measuring any gravitational waves) without any disturbance to determine which slit the particle is going through. This would prevent the particle being put in superposition to form an interference pattern. However, there is a hidden assumption in this argument, as Figure \ref{fig:twogaussians} makes clear. \cite{oppenheim2018post} %newedit
}
	\label{fig:doubleslit}
\end{figure}

What a difference seventy years makes. 
Today, it would be challenging to find a researcher who seriously considers the possibility of not quantising the gravitational field.
There is a clear line, albeit not the only one, between DeWitt's essay, and
the nearly unanimous contemporary view that we must quantize gravity\cite{dewitt1995interview,DeWitt2011pursuit,kaiser2018price}.
 The founder of the Gravity Research Foundation, Roger Babson showed the essay to Agnew Bahnson, who would go on to fund a number of initiatives, including the famous 1957 Chapel Hill conference\cite{cecile2011role}. It was there,
  that the question of whether to quantize the gravitational field was forcefully debated 
 by DeWitt, Bergmann, Wheeler, Salecker, Rosenfeld, Feynman, Louis Witten (the father of Ed Witten) and others. It was Feynman who gave the strongest argument for quantising the gravitational field, claiming that it would be inconsistent to have a theory in which matter was quantum, but the gravitational field was not. A slight variation of his argument\cite{Feynman:1996kb-note} due to Aharonov\cite{AharonovParadoxes-note} is depicted in Figure  \ref{fig:doubleslit}. 
 
 Imagine a double-slit experiment using a massive particle. Depending on its position, this particle produces a different and unique gravitational field. If we consider this field as classical, we could, in principle, measure it with extreme precision to determine the particle's position without disturbance. This knowledge would disclose which slit the particle went through, thereby preventing an interference pattern from appearing on the screen -- a scenario that contradicts actual observations. This argument holds regardless of whether the gravitational field undergoes measurement or not. The mere correlation with the gravitational field prevents the quantum state from being in a pure state that is a superposition of the two locations.

Another argument against leaving the gravitational field classical, is deployed by DeWitt in his essay, who considers the semi-classical Einstein's equation, where one replaces the quantum stress-energy tensor of matter (an operator), by its expectation value\cite{sato1950attempt,moller1962theories,rosenfeld1963quantization}
\begin{align}
	G^{\mu\nu}
	%\stackrel{?}{=}
	\jono{=}\frac{8\pi G}{c^4}\langle{\hat T}^{\mu\nu}\rangle
	\label{eq:semi}
\end{align}
While this at least equates a c-number on the left hand side, with a c-number on the right hand side, he notes that it would violate one of the most fundamental properties of quantum theory, namely the principle of superposition (or linearity), which either leads to a breakdown of causality or the statistical interpretation of the density matrix\cite{gisin1989stochastic,gisin1990weinberg,polchinski1991weinberg,page1981indirect}.
The identification of classical gravity with Equation \eqref{eq:semi} has become so ingrained in the discourse\cite{duff1980inconsistency,page1981indirect,unruh1984steps,carlip2008quantum} that
virtually no-one takes the proposition of a fundamentally classical space-time metric seriously. 
 The pathological non-linear behaviour of the semi-classical Einstein's equation is reason to reject it as a fundamental theory, but this says nothing about other classical-quantum evolution laws which are not based on expectation values.
% As we shall see, there are consistent methods to couple  quantum and classical degrees of freedom.
  %newedit

In the years since the Chapel Hill conference, no-go theorems
and arguments in favour of quantising gravity have persisted, and even gained momentum\cite{dewitt1962definition,
	eppley1977necessity,caro1999impediments,salcedo1996absence,
	sahoo2004mixing,terno2006inconsistency,salcedo2012statistical,barcelo2012hybrid,marletto2017we,Mari:2015qva,Baym3035,belenchia2018quantum,galley2022no}, despite at times, facing resistance \cite{albers2008measurement,kent2018simple,tilloy2018binding}. 
Whatever the cause, the tide of opinion has dramatically shifted since the 50s. Whereas DeWitt appeared to assign equal likelihood to a quantum vs classical space-time metric, present-day quantum gravity researchers are remarkably confident in the quantum nature of gravity, with leading researchers even offering me odds of 1:5000 on a recent wager\cite{bet}.

Can this shift be justified? Or should we pause a moment to re-examine the path we have taken since Chapel Hill? 
It is the purpose of this essay to explore this possibility, and lay out an alternative direction. Whether the next theory of gravity is one in which matter is quantum, and space-time is classical, I have no idea. We are only ever constructing approximate mathematical models of the world around us, and if we're lucky, each iteration will give way to a better one. 
Perhaps gravity will force us to consider theories far beyond our current frameworks\cite{hardy}. But let us here consider the approximation where space-time is a pre-requisite classical structure on which quantum matter lives and acts upon, and examine as best we can, how reasonable this possibility is. 
\begin{figure}[h]
	\includegraphics[width=.7\textwidth, height=7cm]{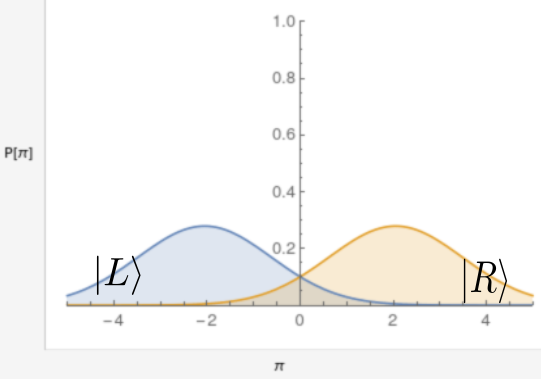}%{twoguassians.png}
	\caption{Consider the Feynman-Aharonov gedanken experiment depicted in Figure \ref{fig:doubleslit}. If in each case of being in state $\ket{L}$ or $\ket{R}$, the particle produces a probability distribution over states of the gravitational field (here represented merely as a probability distribution $P$ over a degree of freedom $\pi$), then the particular state of the gravitational field does not fully determine whether the particle is in the $\ket{L}$ or $\ket{R}$ state, and an interference pattern can result. The broader the probability distribution, the longer the coherence time can be\cite{oppenheim2021gravitationally}.}
	%	In contrast, the local electromagnetic field close to the particle, cannot be measured	At late times, the state the electromagnetic field far from the slits will be slightly different depending on which slit the particle goes through, but these two states, need not be orthogonal. The quantum nature of the electromagnetic field prevents an experimenter from distinguishing these two non-orthogonal states and while the electromagnetic interaction might lead to some decoherence of the particle, if the interaction is small enough the interference pattern remains. On the other hand, if the gravitational field is classical and the interaction deterministic, then one could imagine measuring the gravitational field close to the particle and during its flight, without disturbing the field to unambiguously determine which path the particle went through. One could therefore never get interference fringes (or the uncertainty principle would be violated). Two classical states cannot be different but non-orthogonal. Actually measuring the gravitational field is not required since classical systems are undisturbed -- even if we can't actually measure the field to the required accuracy\cite{mattingly2006eppley}, it is impossible to write down a pure quantum state of the particle which is correlated with a classical gravitational field determined by the particle's position.}
\label{fig:twogaussians}
\end{figure}

The first point to make, is that 
there is an assumption underlying the arguments against a classical gravitational field -- that the classical field is deterministically produced by quantum matter. If instead, the classical field responds to the quantum system through a stochastic process, the Feynman-Aharanov double-slit experiment would not rule out a classical space-time metric. If the gravitational field is not in a deterministic state, but is left in a probability distribution of possible configurations depending on which slit the particle went through, and if there is an overlap in these two distributions as in Figure \ref{fig:twogaussians}, then measuring the gravitational field will not unambiguously determine which slit the particle went through. One can still put the particle in superposition and expect an interference pattern on the screen\cite{refute}. Two probability densities over classical states can be different and non-orthogonal, just like the quantum electromagnetic field after the particle hits the screen will be left in a different but non-orthogonal state depending on which
slit the particle went through. %, causing only partial decoherence. 
A stochastically produced gravitational degrees of freedom only contains partial information about the location of the particle. Since the gravitational field is so weak, 
a large amount of stochasticity is not required to maintain quantum coherence. This is not the case for the electromagnetic field which would require such a large amount of stochasticity, that the diffusion in the field would be observable\cite{oppenheim2021gravitationally}.%newedit

With these no-go arguments addressed, is it possible to find consistent {\it hybrid} theories?  
While there have been numerous failed attempts, an important insight can be gained from the derivations of the most general form of classical stochastic dynamics—namely, the Fokker-Planck or differential Chapman-Kolmogorov equation\cite{gardiner2004handbook}—and the most general form of quantum dynamics in the context of open quantum systems, which is the Lindblad or GKSL equation\cite{GKS76,Lindblad76}. Given a classical and a quantum system,
the joint state is described by a probability density over phase space, and at each point in phase space, a density matrix over Hilbert space\cite{aleksandrov1981statistical}.
 The dynamics is consistent, provided it is (i) linear in the state space and (ii) preserves the form of the state. The first condition is required in order to respect the statistical interpretation of the density matrix. The second condition, mapping states to states, is satisfied if the map is completely positive and norm preserving. In this way, probabilities are mapped to probabilities.
 
Examples of such dynamics, have been known since the 
mid 90s\cite{blanchard1993interaction,blanchard1995event,diosi1995quantum,alicki2003completely,poulinKITP}. And consistent dynamics where the Newtonian potential is treated classically, while matter is quantum have been proposed based on  measurement and feedback processes\cite{kafri2014classical,tilloy2016sourcing,tilloy2017principle}.
There is then a clear path forward: one can find the most general form of hybrid dynamics \cite{oppenheim2018post,UCLPawula}, and
use them to construct a dynamical theory of classical general relativity coupled to quantum fields using the Hamiltonian formulation\cite{oppenheim2018post}. A manifestly covariant theory of gravity\cite{UCLPISHORT} based on a classical-quantum path integral\cite{UCLPILONG} has recently been proposed, together with Zach Weller-Davies.

 Crucially, these proposed dynamics do not correspond to the semi-classical equations. Because the central object is a combined probability density, the models preserve correlations between the quantum system and the classical system. This is the source of difficulty with the semi-classical equations\cite{oppenheim2018post,UCL2022semi} as is apparent by inspecting Equation \eqref{eq:semi}.
 Taking the expectation value erases all information about correlations between the gravitational degrees of freedom on the left hand side side, and the matter degrees of freedom on the other, whether they be quantum or classical. The exact same difficulties arise in a purely classical theory involving two systems if we allow statistical mixtures of states.
 
 While these models were not initially well known in the gravity or high energy physics community, I believe many of my colleagues now agree that it is not logically necessary to quantize the gravitational field.
However, they will argue that for aesthetic reasons, gravity should be treated like any other field theory. ''What is so special about gravity?" they ask. In response, it can be argued that gravity is unique in that it describes the very fabric of space-time in which all matter exists. Having this classical background structure appears to be essential for a consistent understanding of quantum theory.
While other field theories can be partially recast in a geometric form, they do not give the same geometry that all matter move in -- each charge requires its own geometry. Only the universal nature of gravity, through the equivalence principle, dictates that it alone can be thought of as giving the background space-time in which we perform all our experiments.
If we take the geometric nature of gravity seriously, then we should be open to the possible that gravity is not mediated by gauge bosons as other forces are, but instead, by geometry which may be of a different nature. 
 More to the point, gravity's dominant contribution is simply that clocks run at different speeds in different regions of space. 

Is it possible to have the speed of this unequal flow of time from place to place, be a quantum observable, given that quantum field theory describes the changing state of all fields in relation to a classical parameter time $t$? Maybe. But even posing the questions that physicists typically ask, is a dubious proposition.
%This background classical time, is an essential ingredient of standard quantum theory. 
If we are trying to solve some initial value problem, then we specify a state of quantum fields at an initial time, and describe how this evolves. But if the quantum field is the metric itself and it can be in superposition, then even specifying that there is an initial space-like hyper-surface could be problematic. This is almost certainly specifying information which is complementary to conjugate degrees of freedom of the spatial metric. We can certainly go ahead, choose some initial slice, and specify initial data including the spatial metric on it. We can work in the Hamiltonian framework of general relativity, and evolve this initial quantum state forward provided we choose some folation of space-time into a one-parameter family of space-like hypersurfaces $t$. But the resulting quantum theory is likely to depend on this choice. We appear to require some classical background metric, and it's only the fluctuations around this classical metric which we treat as quantum degrees of freedom. Does the resulting quantum theory depend on the choice of classical background that we perturb around? It's hard to imagine it doesn't.
%
%Or take the usual specification of the commutators of two fields at different space-time points. If the points are space-like separated, we expect the commutator to vanish. But as Carlip has emphasised, whether the two points are space-like separated depends on the state of the gravitational field. A  no longer a kinematic property

This leads to just one facet of the famous {\it problem of time}, or more accurately, one of many problems of time\cite{kuchar1992time,kuchavr2011time,isham1993canonical,anderson2012problem}.
In classical general relativity, the choice of $3+1$ splitting does not break general covariance provided the {\it constraints} are satisfied, with the Hamiltonian density $\mathcal{H}$, and momentum density $\mathcal{H}^a$, constrained to be zero at each point $x$.  While this still allows for non-trivial evolution in the classical case, the quantum version of imposing the Hamiltonian constraint
on the state $\ket{\psi}$ is the Wheeler-DeWitt equation 
\begin{align}
	\hat{\mathcal{H}}\ket{\psi}\approx 0
\end{align}
with  $\mathcal{H}$ promoted to an operator. Since the Hamiltonian is an integral of pure constraints, this dictates that the quantum state is frozen, with states of well-defined energy failing to evolve. Attempts have been made to circumvent this, such as considering relational degrees of freedom with respect to a physical clock\cite{rovelli1991time,goeller2022diffeomorphism}, but there is generally no physical system which acts like a good clock\cite{unruh-wald-timebackwards}.

The difficulties associated with time and the Hamiltonian constraint, were never overcome by DeWitt and the canonical quantisation approach. %\todo{ref}. 
Nor were they overcome by its successor, Loop Quantum Gravity\cite{sen1982gravity,ashtekar1986new,ashtekar2004background,perez2004introduction,nicolai2005loop,smolin2006case,thiemann2007loop,rovelli2008loop,thiemann2003lectures,rovelli2004quantum,ashtekar2021short}, which would eventually be forced to move away from the Hamiltonian formulation of general relativity, and towards a path-integral approach where the fundamental degrees of freedom are not the spatial metric, or loop degrees of freedom, but {\it spinfoams}. Whether the classical, low energy limit of spinfoam models recover Einstein's equations, or even space-time itself is far from clear. In part because it's hard to see how a path integral can give the transition amplitudes between two times, when time itself is meant to be merely an emergent property. 

String theory\cite{veneziano1968construction,PhysRevD.46.5467}, by far the dominant approach to reconciling gravity with quantum theory, fares no better when it comes to these conceptual issues. Or rather, it side-steps them by considering strings moving in a particular classical background space-time. %Background independent approaches such as {\it String Field Theory}\cite{PhysRevD.46.5467} are in their infancy.
One can instead try to consider space-times which have a boundary, and describe the evolution of diffeomorphism invariant observables in terms of boundary time as is done via the AdS/CFT conjecture\cite{Maldacena:1997re,Witten:1998qj}. While this doesn't describe our universe, it does provide a candidate quantum theory in the case of Anti-deSitter space (AdS). But again, we are required to presuppose a classical and well defined time slicing at the boundary, in which the Conformal Field (CFT) evolves. 

While boundary conditions are naturally part of the specification of a theory, here the boundary conditions, that the metric be conformally flat, are specified classically. In the quantum case, we don't expect components of the metric to commute as observables. In AdS/CFT, the boundary acts similarly to a reference frame, with a well-defined time, relative to which bulk observables can be defined. Indeed in order to define background independent observables without a boundary, one appears to be pushed towards defining observables relative to the worldline of an observer\cite{witten2023background}, bringing with this all the baggage and difficulties faced in the canonical quantisation approach with relational degrees of freedom. 

It is also worth remarking, that although AdS/CFT is widely regarded as defining a quantum theory of gravity (albeit in asymptotically AdS universes), the precise nature of the duality is still a work in progress, especially in the case of evaporating black-holes. Evidence is mounting that the duality should not be regarded as an isometry between the boundary CFT theory and the bulk\cite{akers2022black}, and there is disagreement on whether the bulk is dual to a single boundary CFT or an ensemble of CFTs\cite{marolf2020transcending}, an issue related to the presence of replica wormholes in the path integral\cite{penington2022replica,almheiri2020replica} and {\it the factorisation problem} in AdS/CFT\cite{maldacena2004wormholes}. While the CFT is clearly a consistent quantum theory, whether it consistently describes an evaporating black hole, has not in my view been established, although there is confidence that this is the case.

It is often said in the context of quantum gravity, that space-time is an emergent phenomenon. In the case of AdS/CFT, this can be made more precise, in the sense that space-time in the bulk emerges from entanglement in the boundary CFT\cite{van2010building,maldacena2013cool}. But on the boundary, we need to imagine a classical space-time, because entanglement is a property of distant regions at a single time $t$. Entanglement between where and where exactly?  And evolving with respect to which time coordinate? We are currently unable to conceive of quantum theory without the crutch of a classical space-time geometry. While it may be that space-time emerges from some other degrees of freedom, we only know how to describe these other degrees of freedom as living on some other background space-time.
It strikes me as implausible that space-time should emerge from a quantum theory which is merely set in some different background space-time geometry. It feels too much to ask of the universe, that a theory in which space-time is merely emergent, can be described by that exact same theory merely transposed to a different non-emergent space-time.%  And since quantum theory as we currently know it, requires a background geometry, it does not seem plausible that space-time should emerge from a quantum theory in which we must first presuppose a different space-time geometry.

That is not to say that there isn't some intriguing evidence for string theory -- it gives the correct value for the black-hole entropy for a class of extremal black holes\cite{strominger1996microscopic}. But throw in the need for supersymmetry and extra-dimensions, both of which are unobserved at current energy scales, and it's hard not to feel the need to hedge our bets, in case we're heading in the wrong direction. Now, my colleagues are struggling valiantly to overcome these challenges.  We are all flailing about as best we can. It's certainly conceivable that they, and those who follow in their footsteps, will find a solution to the problem of time and the other conceptual issues that quantum gravity faces. But it's hard to envisage what the solution will even look like. This is a conceptual issue which strikes at the very heart of what it means to quantize space-time geometry. We may have to accept that quantum theory simply requires a classical flow of time.  
A universe in which space-time is classical while matter is quantum may seem improbable, but the idea of a universe in which our classical understanding of space-time is completely disrupted, while quantum theory remains unchanged seems as unlikely.

While we are nearing the centenary of our failure to quantize gravity\cite{rosenfeld1930quantelung,rosenfeld1930gravitationswirkungen}, 
constructing a consistent theory in which space-time is classical is not without its challenges either: the path integral which can be shown to give a completely positive map only gives the trace of Einstein's equations in the classical limit, while the path integral which gives the full Einstein's equation has not been proven to be a completely positive norm-preserving map\cite{UCLPISHORT}. One also has to show that the theory is regularisable, and isn't ruled out by constraints on diffusion and decoherence\cite{oppenheim2021gravitationally}, or heating effects\cite{bps,gross1984quantum}. 
%newedit

If successful, the project would have at least two additional significant implications. First, since stochasticity is required, it would lead to the conclusion that information may be fundamentally destroyed. Although this is controversial, I find it particularly compelling in light of the black hole information problem\cite{hawking-bhinfoloss,hawking-unpredictability,preskill-infoloss-note}, and its sharper version, the AMPS paradox \cite{almheiri2013black,braunstein2009entangled}. %newedit
While it is possible that a purely quantum theory could somehow allow for fundamental information destruction, there is a no-go argument due to Coleman, who argued that they merely result in false decoherence corresponding to unknown coupling constants\cite{coleman1988black}.  Coleman's argument however, does not apply to classical-quantum theories\cite{oppenheim2018post}.

Of course, the idea that black-holes destroy information may be even more heretical than the idea that gravity should not be quantized. Here again, this was once considered an open question, while today, you'll have a tough time finding researchers who believe that gravity causes a break-down in predictability. But here as well, opinions on this matter are more questions of taste rather than based on scientific evidence. While there are occasionally claims that the information paradox has been solved\cite{penington2022replica,almheiri2020replica,musser2020most}, invariably there is little agreement on the solution.% and the black-hole information problem gets solved as often as the seasons change. 
The AMPS paradox warns us that any attempt to cling to a deterministic theory requires sacrificing  the equivalence principle or low energy effective field theory, and we have as strong a reason to believe in these, as the alternative -- a breakdown in predictability\cite{CaseForNonUnitarity}.

This breakdown in predictability is related to the second additional consequence: the mysterious measurement postulate and Born's rule of quantum theory would not be needed\cite{GisinMeas,UCL2022semi,UCLBORNRULE}. The quantum degrees of freedom inherit some classicality from space-time, which cause the quantum system to localise. In a rough sense, this appears to be similar to the gravitationally induced decoherence of the wave-function, conjectured by Karolyhazy, Diosi and Penrose\cite{karolyhazy1966gravitation,diosi1989models,penrose1996gravity}. This is in part because a classical field interacting with a classical one, necessarily induces  decoherence\cite{hall2005interacting,tilloy2016sourcing,tilloy2017principle,PoulinPC}. However,  Isaac Layton, Zach Weller Davies, and myself, found that one can always choose  quantum-classical dynamics, such that the quantum state remains pure when we condition over the classical trajectory (in this case, space-time itself)\cite{UCL2022semi}. There is a breakdown in predictability of the classical degrees of freedom which corresponds to the probabilities associated with quantum theory. But because the quantum state is pure conditioned on the trajectory of the classical system, the decoherence is merely due to integrating out the classical system\cite{UCLcoherence}.

This brings us to another reason to be cautious of quantum gravity. Quantum theory was developed to describe the results of measurement outcomes performed in the lab by an experimenter who is external to the system. She associates a quantum state to the physical system of interest, but it is not just her clock that is treated classically -- so too are her measuring apparatus, the notebook she records her observations in, her lab assistants, and herself. Applying the quantum formalism to describe the  wave-function of the universe, as is done in quantum gravity (or quantum cosmology), is certainly a reasonable thing to try, but if successful, it would be an extension of physical law from one domain to another, the likes of which we have rarely seen. 

Returning to the prevailing view among physicists of DeWitt's time, the prime reason for opposing the attempt to quantize of gravity, was ''the experimental fact that gravitation has never been observed to take part in physical events on a quantum level, and where there is no evidence it is bad form to speculate.''\cite{dewitt1953new} seventy years later we are in a different position. A number of experiments have been proposed to test the quantum nature of gravity based on detecting gravitationally induced entanglement \cite{kafri2013noise,bose2017spin,marletto2017gravitationally,belenchia2018quantum,carney2021using,carney2022newton,overstreet2022inference}, or ''quantumness''\cite{lami2023testing}. These experiments may be decades away and will require significant effort. But current observations based on Cavendish experiments and coherence times can already be used to rule out part of the parameter space of theories in which space-time is treated classically\cite{oppenheim2021gravitationally}. 

While the lack of feasible experiments was viewed as a reason not to quantize gravity,  today's promise of feasible experiments could be viewed as a reason to pursue alternatives. 
Perhaps we will succeed in directly quantizing space-time geometry, or perhaps it will be found to be an emergent property of some higher quantum theory, such as spinfoams, strings, or entanglement. But we should also consider the possibility that somewhere along our journey, whether at Chapel Hill or elsewhere, we
took a wrong turn. And now is as good a time as any, to re-examine our belief that gravity must have a quantum nature. 
 We can speculate all we want, but fundamentally, we must turn to experiment to settle the question.

\newpage
%\bibliographystyle{apsrev4-1}
%\bibliography{rethink,common/refgrav2,common/refcq,common/refjono2,common/rmp12,common/pqg_v3_extras,common/NewtLimitbib}
%\appendix

\end{document}